\documentclass[prl,twocolumn,showpacs,amsmath,amssymb]{revtex4}
\usepackage{graphicx}

\begin{document}

\preprint{Preprint}

\title{Observation of a Triangular to Square Flux Lattice 
Phase Transition in YBa$_2$Cu$_3$O$_7$}

\author{S.P. Brown$^{1,\dagger}$,
D. Charalambous$^1$,
E.C. Jones$^1$,
E.M. Forgan$^1$,
A. Erb$^2$,
J. Kohlbrecher$^1$}

\affiliation{$^{1}$School of Physics and Astronomy, University of Birmingham, Birmingham B15 2TT, U.K.\\ 
$^{2}$ Walther-Meissner Institut, Garching D-85748, Germany.\\
$^{3}$Paul Scherrer Institut, Villigen PSI, CH 5232, Switzerland.
}
\date{\today}

\begin{abstract}
We have used the technique of small-angle neutron scattering to observe 
magnetic flux lines  directly in an 
YBa$_2$Cu$_3$O$_{7}$ single crystal at fields higher than previously reported.
For field directions close to perpendicular to the CuO$_2$ planes,
we find that the flux lattice structure changes smoothly from a distorted 
triangular co-ordination to nearly perfectly 
square as the magnetic induction approaches $11\thinspace\rm{T}$. The 
orientation of the square flux lattice is as expected 
from recent {\it d}-wave theories, but is $45^\circ$ from 
that recently observed in La$_{1.83}$Sr$_{0.17}$CuO$_{4+\delta}$.
\end{abstract}


\maketitle

The technique of small-angle neutron scattering (SANS) from 
flux lines has a long and honorable record 
in measuring the properties of flux lines in 
superconductors. However, it continues to bring new 
dividends, especially in unconventional superconductors, 
since important information about the nature of 
the superconducting state is often revealed by the flux line 
lattice (FLL) structure, for example 
~\cite{SROKealey,LSCOGilardi,UPt3Huxley,BcarbDon}. 
The diffraction pattern 
not only reveals the co-ordination and perfection of the 
FLL, and its correlation with the crystal lattice, but also 
the absolute intensity may be used to determine the actual 
spatial variation of the magnetic field within the mixed 
state and the values of the coherence length and penetration 
depth~\cite{SROKealey,NbForgan,SROTanya}. In the simplest 
approximation, flux lines would order in a regular 
triangular FLL; however, anisotropy of the Fermi surface
or of the superconducting order 
parameter can cause distortions of the triangular lattice 
or transitions to other structures. The simplest situation 
in a high-$\kappa$ material is anisotropy of the magnetic 
penetration depth associated with effective mass anisotropy 
\cite{Thiemann}; for example, the anisotropy in the {\it ab} plane 
of YBa$_2$Cu$_3$O$_{7-\delta}$ (YBCO) leads at low values of field to a corresponding 
distortion of triangular FLLs~\cite{Johnson}. At the lower 
values of $\kappa$ in borocarbides, ``nonlocal'' effects 
are expected and observed \cite{KoganLondon,dewilde,BcarbDon} 
to give a variety of FLL distortions and 
transitions. If the superconducting order parameter has a 
different symmetry from that of the crystal, this can again 
be revealed via its effects on the FLL structure, for 
instance in the {\it p}-wave superconductor Sr$_2$RuO$_4$~\cite{Agterberg,SROTanya}. In general in {\it d}-wave superconductors~\cite{ichioka,shiraishi}, 
there is expected to be a tendency 
towards a square FLL as the field is increased and the 
anisotropic flux line cores overlap. This may be the cause 
of the FLL phase transition recently observed in overdoped  
La$_{1.83}$Sr$_{0.17}$CuO$_{4+\delta}$ (LSCO) at 
the comparatively low field of $0.4\thinspace\rm{T}$~\cite{LSCOGilardi}. 
According to~\cite{ichioka}, the FLL 
nearest-neighbour directions should lie along the directions 
of the nodes of the order parameter, which 
would be at $45^\circ$ to the Cu-O bonds in the 
superconducting layers. This is not, however, the 
orientation of the square FLL observed in LSCO~\cite{LSCOGilardi}. 
The {\it orientation} may instead 
be controlled by band structure effects~\cite{MachidaBS}, 
even if the {\it symmetry} of the FLL is 
controlled by {\it d}-wave effects. It has been suggested 
that an peak effect in magnetisation 
measurements on overdoped YBCO may be a signature 
of a continuous triangular-to-square FLL transition in this 
material at high fields \cite{Knigavko}. However, others have suggested
 that there is a glass transition in this region \cite{Nishizaki}.
Only by {\it direct} measurements may such suggestions be tested and the 
correlation between FLL and crystal lattice (or 
superconducting order parameter) determined.

Our experiments were performed on the SANS-I instrument at  
SINQ, PSI, 
Switzerland. Cold neutrons (8$\thinspace${\AA} to $14\thinspace${\AA}, with a FWHM wavelength 
spread of $10\thinspace\%$) were collimated over distances from $4.5\thinspace\rm{m}$
to $15\thinspace\rm{m}$, depending on the field and hence $q$-range required. The 
diffracted neutrons were registered on a 
$128\times128\times7.5\thinspace\rm{mm}^2$ multidetector, which was 
similarly adjustable in distance from the sample. The 
undiffracted main beam was intercepted by a cadmium 
beamstop. A magnetic field of up to $11\thinspace\rm{T}$ applied 
approximately parallel to the neutron beam, was provided 
by a cryomagnet with a field 
uniformity of $0.2\thinspace\%$ over a $1\thinspace\rm{cm}$ sphere.
 A variable temperature insert containing He heat exchange 
gas allowed sample temperatures from $1.5\thinspace\rm{K}$ to 
$300\thinspace\rm{K}$. 
The sample was a 40$\thinspace{\rm mg}$ low--twin-density high-purity single crystal of 
YBCO grown in a BaZrO$_3$ crucible~\cite{erblovesbazro} and oxygenated close to $\rm{O}_7$ by high-pressure oxygen 
treatment in order to reduce pinning by oxygen vacancies in 
the Cu-O chains~\cite{Erbprep}. It was 
therefore overdoped and had a $T_{\rm c}$ of $86\thinspace\rm{K}$.
It was initially mounted with its ${\bf c}$-axis parallel to the 
field direction.
In order to satisfy 
the Bragg condition for each diffraction spot in turn and 
hence establish the FLL structure, the cryomagnet 
and sample together could be rotated or tilted to bring the 
FLL Bragg planes to the appropriate small angles ($\sim 1^\circ$) to the 
incident neutron beam.
\begin{figure}[t]
\includegraphics[width = 7 cm]{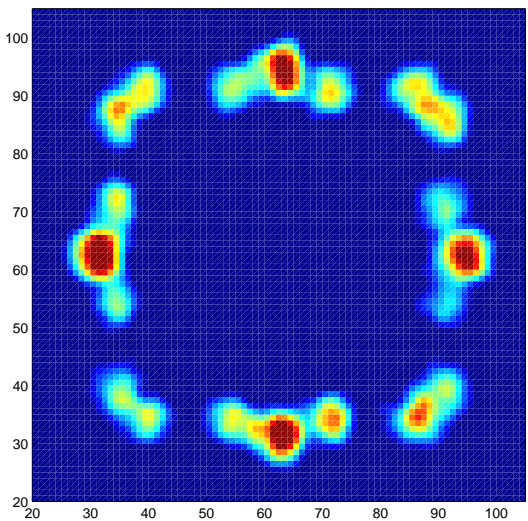}
\caption{FLL diffraction pattern at 1 T. The figure shows 
the counts on the SANS multidetector at $4\thinspace{\rm K}$ (minus 
backgrounds obtained above $T_{\rm c}$) summed over a range of 
angles between the field direction and the neutron beam. 
Noise at the centre of the picture has been masked. The 
cryostat was rocked by $\pm 1^\circ$ about horizontal 
and vertical axes, ensuring that all spots in the 
diffraction pattern from the sample are detected. The $\{110\}$ 
directions, corresponding to twin plane directions are 
vertical and horizontal in this picture. In all cases, 
the FLL was formed by applying the field above $T_{\rm c}$ and 
cooling.}
\label{fig1}
\end{figure}

In Fig.~1 is shown the FLL diffraction pattern obtained at 
the low field of $1\thinspace\rm{T}$. The most obvious feature 
of this pattern is its fourfold symmetry which reflects the 
average fourfold symmetry of the twinned orthorhombic
structure of our YBCO sample. However, the FLL structure 
itself has {\it triangular} coordination, and the symmetry 
of Fig.~1 arises from four orientations of distorted 
triangular FLLs, present in different domains in the sample, 
as was first observed by Keimer{\it et al.}~\cite{Keimer}. 
The diffraction spots arising from these four triangular 
lattices are represented in Fig.~2. It appears that the {\it 
distortion} of the individual lattices arises mainly from the {\it 
a/b} anisotropy present in each orthorhombic domain in the 
crystal~\cite{ForganLee,Walker}. This interpretation was 
confirmed by measurements on an untwinned sample~\cite{Johnson} 
which show diffraction spots distributed around an {\it 
ellipse} aligned with the ${\bf a}$ and ${\bf b}$ axes. The 
ratio of the principal axes of the ellipse should represent the 
anisotropy of the London penetration depth for $B_{\rm c1}\ll B \ll B_{\rm c2}$
~\cite{Thiemann}. The value we observe for the anisotropy 
ratio, $\gamma_{ab}$, in our sample is $1.28(1)$, whereas many
estimates of this quantity are rather larger \cite{largerabanisotropyrefs}. However comparable values to ours were obtained by measurements on a separate
 untwinned sample using neutrons \cite{Johnson} and muons and torque magnetometry \cite{leetorque}. Our results are also corroborated by recent 
surface-sensitive measurements
using a novel atomic-beam magnetic-resonance technique \cite{magres}.
It seems likely that the precise value of $\gamma_{ab}$ depends on the degree of
perfection of the Cu-O chains along the ${\bf b}$ direction
\cite{Tallon_anisotropyref}. The 
{\it orientation} of the triangular FLLs has been 
ascribed to pinning of a pair of spots, and hence planes of 
flux lines, to the twin planes~\cite{ForganLee,Walker}. 
However, results reported later in this Letter also support the 
existence of a correlation between the nearest-neighbour FLL 
directions and the directions of 
zeroes of the {\it d}-wave order parameter.
\begin{figure}[t]
\begin{center}
\includegraphics {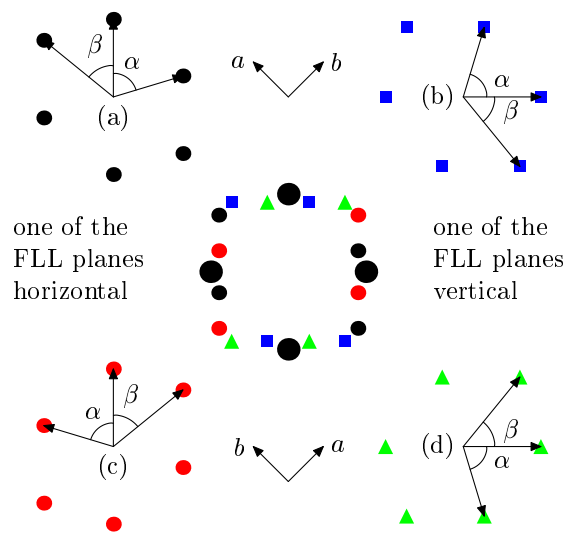}
\end{center}
\caption{Schematic of diffraction patterns from four 
distorted triangular FLLs that together account for the 
pattern in Fig.~1. In each orthorhombic domain in the 
twinned crystal, there are two orientations of FLL
(eg. (a) and (b)), derived 
by taking a regular hexagonal pattern and distorting it by 
the {\it a/b} anisotropy \protect\cite{Thiemann}. 
The more intense pair of spots in each pattern is aligned 
with one of the $\{110\}$ directions. The centre figure is the
superposition of the four FLL domains (a), (b), (c) and (d), which
gives rise to the pattern in Fig.~1.
The angles between reciprocal 
lattice vectors, $\alpha$
and $\beta$, are defined for use in Fig.~5} 
\end{figure}

In Fig.~3, we show diffraction patterns taken at higher 
fields.
The data taken at $7\thinspace\rm{T}$ show 
a distortion of the triangular FLLs so 
that some of the weaker spots are closer to the strong 
spots, and others have moved towards the 
diagonals. There is clearly another source of 
distortion than pure {\it a/b} anisotropy. Finally at $11\thinspace\rm{T}$, 
the FLL has become almost exactly
square, with the weak corner spots now playing 
the role of second order $\{1,1\}$ spots of a 
square FLL instead of first order spots from a distorted 
triangular FLL. 
\begin{figure}[t]
(a)\includegraphics[width = 5.8cm]{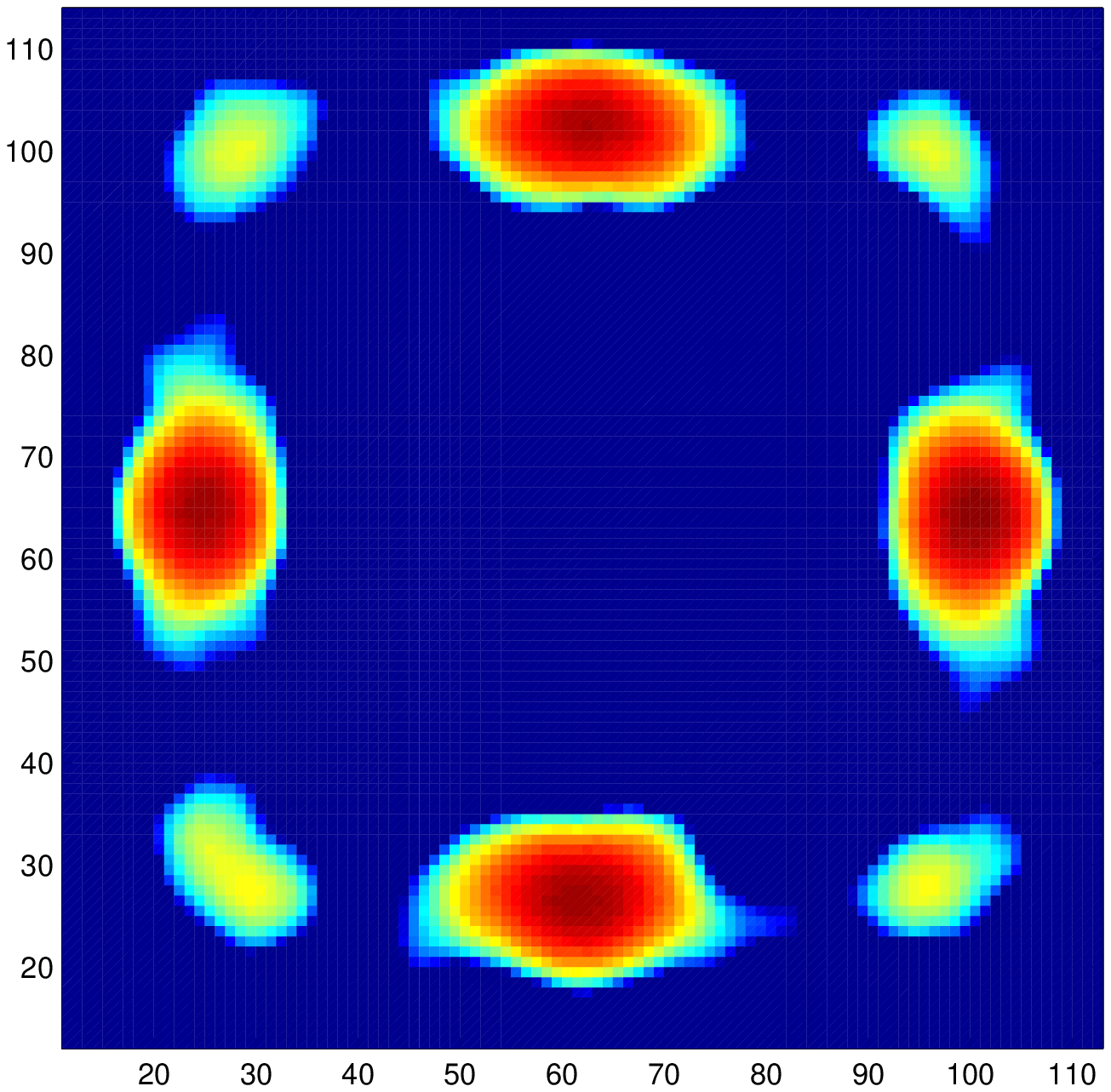}
(b)\includegraphics[width = 5.8cm]{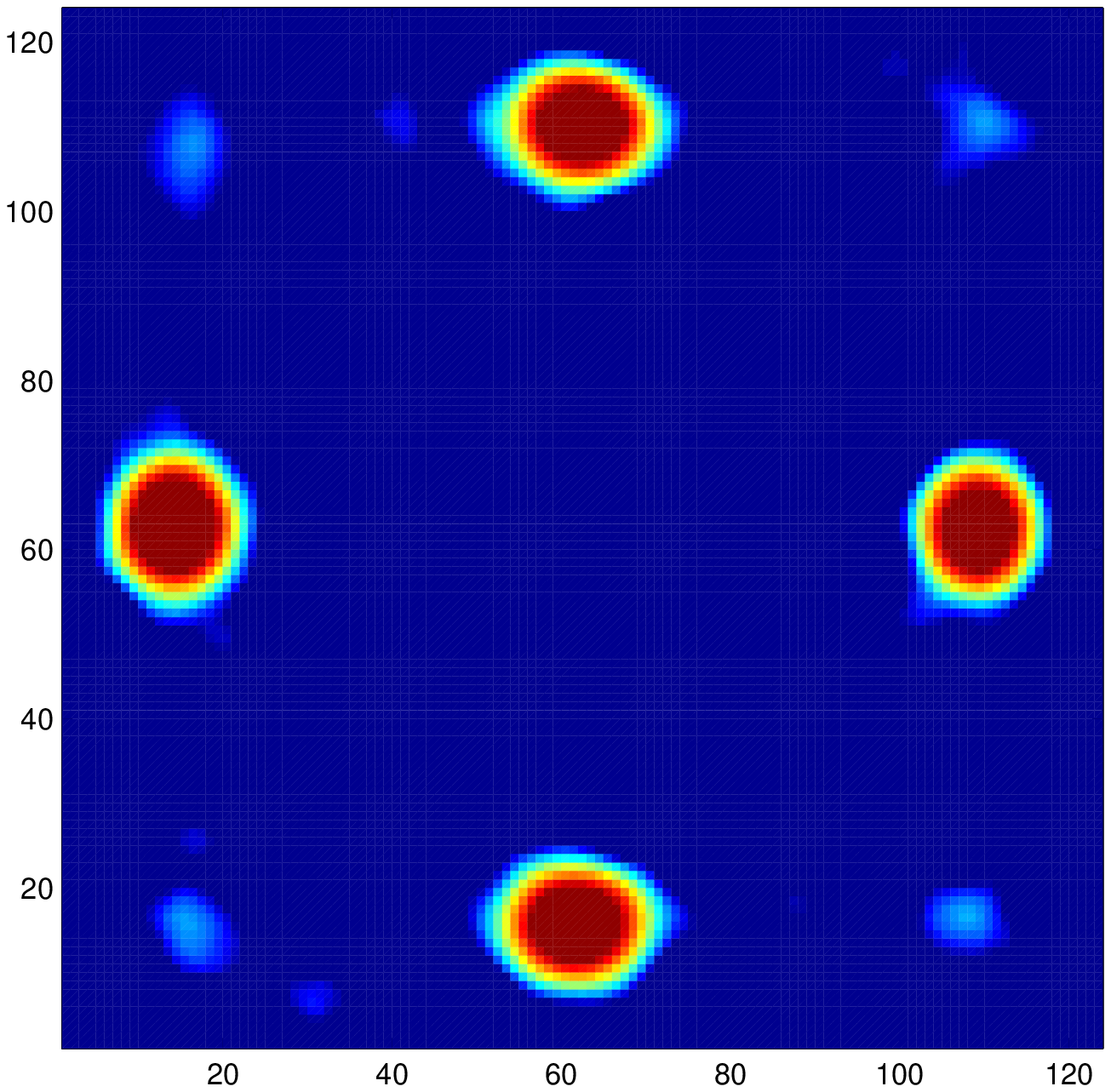}
\caption{FLL diffraction patterns, as in Fig. 1 (but with a logarithmic
intensity scale and smoothed to make the weaker spots more clearly visible), (a) at 
$B=7\thinspace\rm{T}$, (b) at $B=11\thinspace\rm{T}$, 
showing the change in position of the weaker spots as 
the field is increased}
\end{figure}
In order to investigate this steady change in the FLL 
structure with field, we rotated the crystal about the 
vertical axis in Fig.~3, so that the field was $5^\circ$ 
from the ${\bf c}$-axis. This was done in order to break the 
degeneracy between those FLL structures giving strong 
vertical diffraction spots~(Fig.~2(a) and (c)), 
and those giving strong horizontal spots~(Fig.~2(b)~and~(d)). 
Within anisotropic London theory, this small angle of rotation should
make a negligible change to the FLL distortion.
As shown in Fig.~4., we found that at high fields the 
FLL structures giving horizontal spots were 
suppressed and instead only the structures 
depicted in Fig.~2(a) and (c) were observed. 
The advantage of this arrangement is that the pairs of 
spots {\it near} the horizontal axis in Fig.~4 could be 
observed easily, without being overlaid by the strong ones {\it on}
the axis. This allowed us to measure accurately the spot positions
and hence the FLL distortion.
Nevertheless, at high field, this pair of spots overlaps,
but by assuming that the spot size is independent of field, we may 
estimate the angle between them even when they overlap.
Further measurements of spot positions allow us to give a complete 
description of the FLL distortions versus field in terms of the angles
between the FLL reciprocal lattice vectors. The results of this analysis
are shown in Fig.~5. It is clear that the low field 
structure progressively changes with increase of field, although in our 
available field range the FLL never exactly 
reaches a perfectly square shape. This may partly be because the
phase transition is at the extreme of our available field range,
but is also clearly a result of the orthorhombic structure of YBCO. 
As depicted in the inset to Fig.~5, the {\it a/b} anisotropy of 
each domain must, on symmetry grounds, distort a square lattice 
to a rectangular one, causing a slight splitting of the ``square'' spots from a twinned
crystal such as ours.
\begin{figure}[t]
(a)\includegraphics[width = 5.8cm]{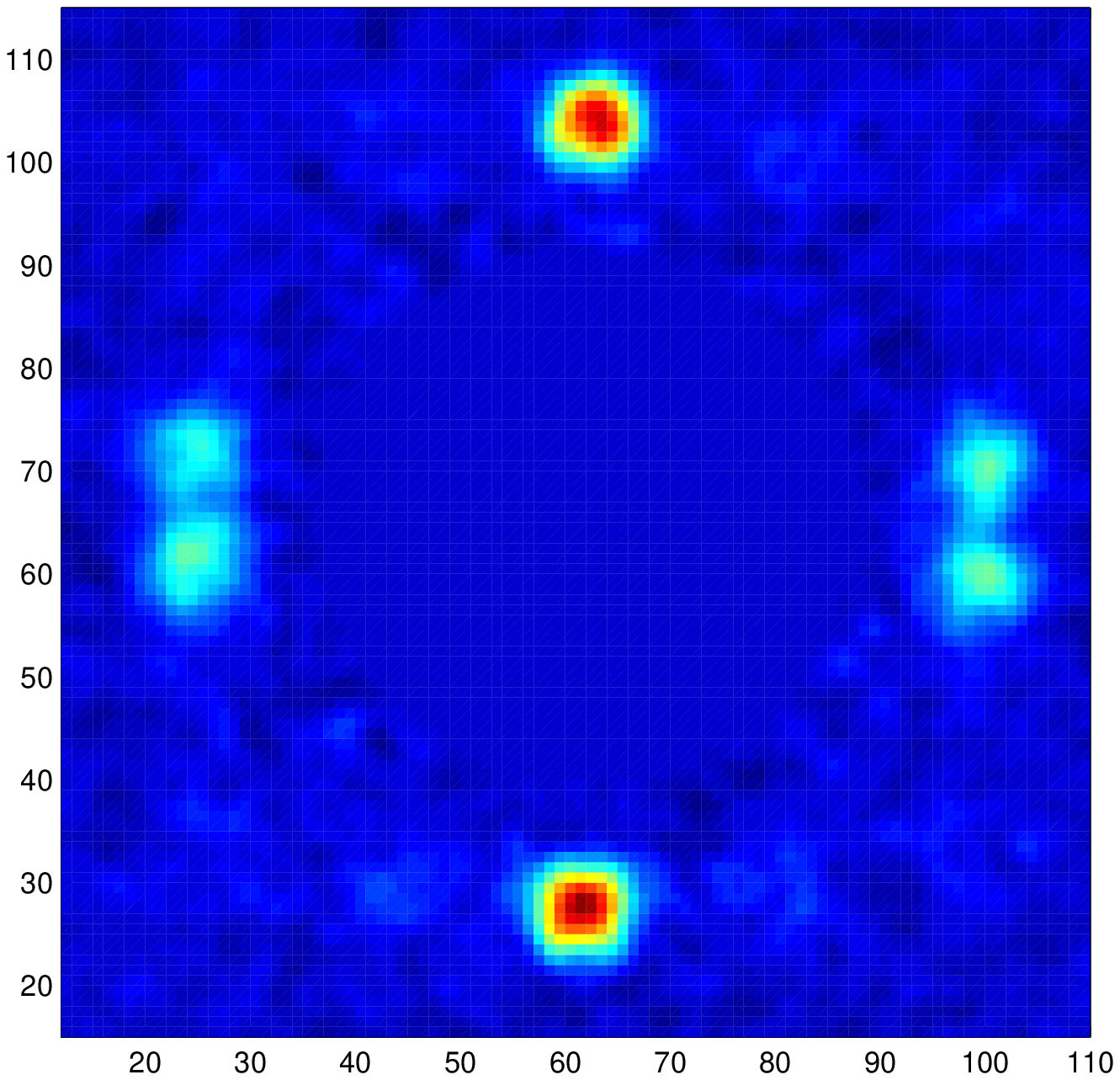}
(b)\includegraphics[width = 5.8cm]{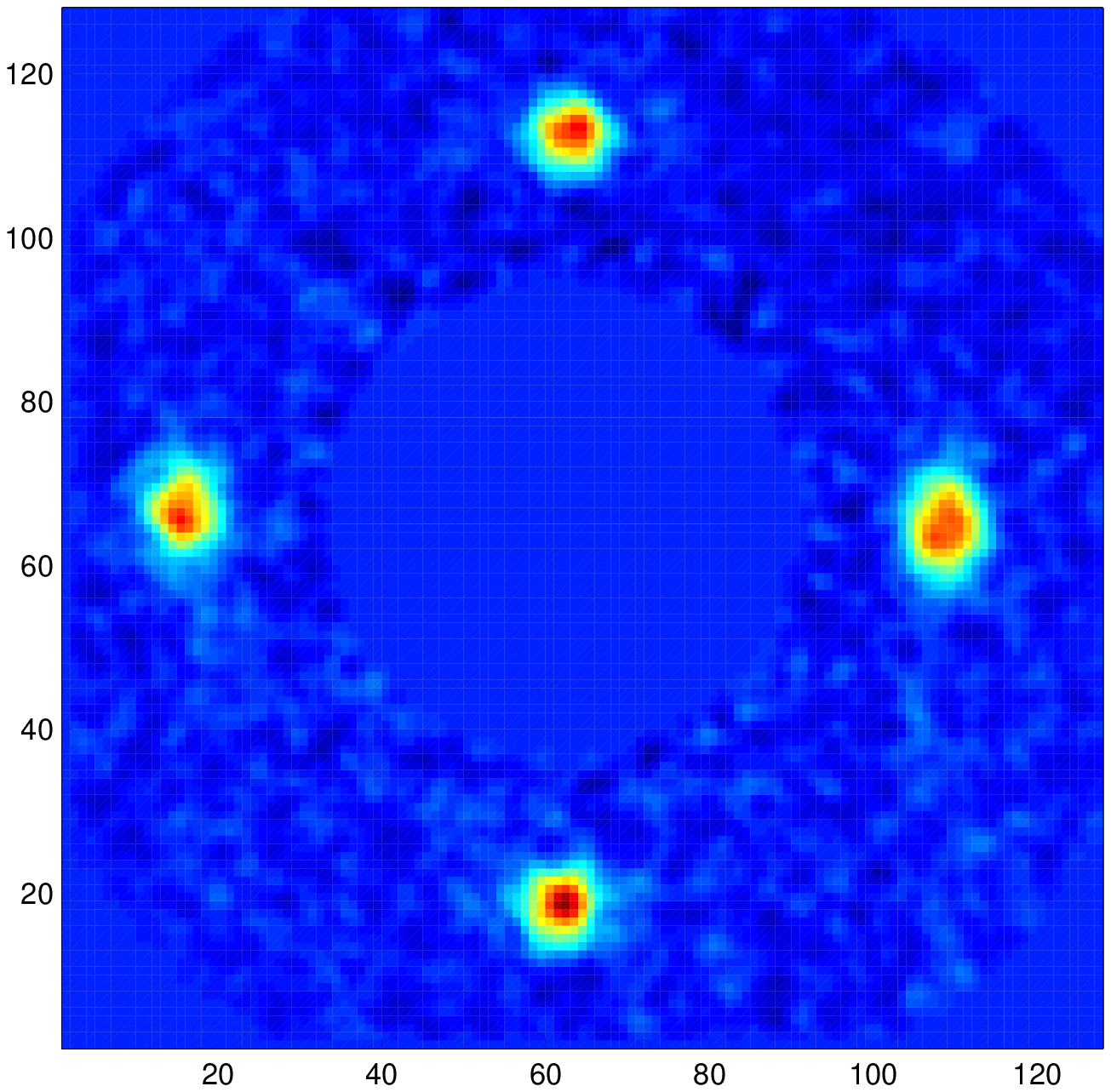}
\caption{FLL diffraction patterns with the field rotated 
5$^\circ$ from the ${\bf c}$-axis, to give only two FLL domains: (a) 
at $B=7\thinspace\rm{T}$, (b) at $B=10.8\thinspace\rm{T}$}
\label{fig4}
\end{figure}

We have further investigated \cite{future} the temperature-dependence of the FLL distortion shown in fig.~5. We find that with increasing temperature, the FLL structure changes more towards triangular. Thus the boundary between square and triangular phases must curve up in field as temperature is increased. We would expect this if the triangular to square transition is due to $d$-wave effects, as the nature of the pairing becomes less important as $k_{\rm B}T$ becomes comparable with the magnitude of the gap. The shape of the phase boundary is similar to that seen in an overdoped sample by macroscopic measurements \cite{Nishizaki}, but not the same as that proposed in ref. \cite{Knigavko}. 
Unlike LSCO \cite{LSCOGilardi}, the {\it orientation} of the FLL that we observe is aligned as expected from $d$-wave theories \cite{ichioka,shiraishi}. It may be argued that twin planes, which are present in LSCO and YBCO are controlling the FLL orientation. To rule this out, measurements were also taken with the field at an angle to both sets of twin planes in our sample and the shape and orientation of the FLL was essentially unchanged. One should also note that the predicted difference in free energy between the two orientations of a square FLL is much larger than that between any triangular and the lower energy square orientation \cite{ichioka}. We also note that a similar correlation between FLL orientation and probable direction of $d$-wave nodes has recently also been observed in CeCoIn$_5$ \cite{CeCoInref}. Further support of the $d$-wave origin of the triangular-square transition in YBCO is the value of the transition field, which is a similar order of magnitude to the predicted $0.15B_{\rm c2}$ \cite{ichioka}. We do not believe that in a large-$\kappa$ material such as YBCO, the transition is due to nonlocal effects as has been proposed for borocarbides \cite{dewilde}.
We note that our {\it bulk} observations of FLL structure are not in complete agreement with surface measurements by STM techniques \cite{fischer,shibatastm}. It is possible that the FLL adopts a different structure near the surface, or that the surface has different doping from the bulk. It is clear \cite{Nishizaki} that the phase diagram is strongly doping-dependent.
%
\begin{figure}[t]
\begin{center}
\includegraphics[width=8cm]{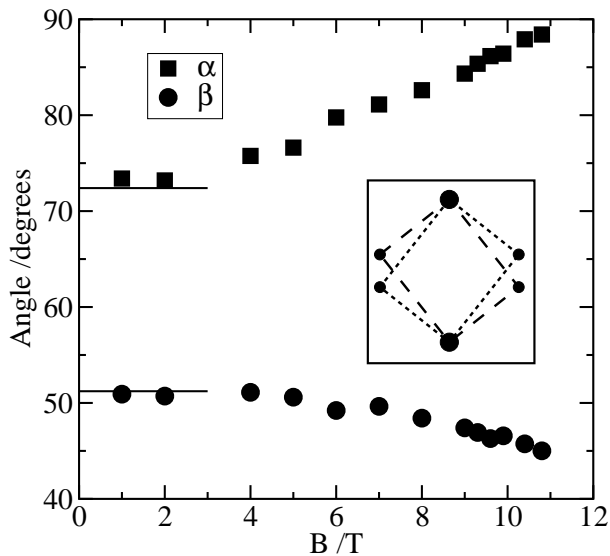}
\end{center}
\caption{Field-variation at $5\thinspace\rm{K}$ of two of the angles 
between reciprocal lattice vectors, $\alpha$ and $\beta$, depicted in
Fig.~2. (Errors are comparable with the marker size.) All data were obtained with $\bf{B}$ parallel to the crystal ${\bf c}$-axis
except the data for $\alpha$ at fields greater than $6\thinspace\rm{T}$,
which were taken with $\bf{B}$ at $5^\circ$ to ${\bf c}$ in order to resolve this
angle more clearly (see text).
A {\it regular} hexagonal lattice 
would have $\alpha = \beta = 60^\circ$, and an exactly square one 
$\alpha =90^\circ$ and $\beta = 45^\circ$.
Also marked by horizontal lines are predictions of anisotropic London theory 
\protect\cite{Thiemann} for $\alpha$ and $\beta$, using a basal plane 
anisotropy, $\gamma_{ab} = 1.28$, and assuming that one pair of spots is
tied to the $\{110\}$ directions.
In the inset is shown the orthorhombic distortion from an exactly square pattern (exaggerated for clarity), expected in the two orthorhombic domains present in our crystal.
}
\label{fig5}
\end{figure}

In conclusion, using small-angle neutron scattering, we have directly 
observed a change from triangular to square co-ordination of 
the flux line lattice as a function of magnetic field 
in fully oxygenated YBa$_2$Cu$_3$O$_7$. This phase transition is
most naturally interpreted as a consequence of the {\it d}-wave
character of the order parameter, which is expected to be more
prominent at high magnetic fields, where the flux line cores
begin to overlap. The {\it orientation} of the FLL is as expected
from an isotropic {\it d}-wave theory~\cite{ichioka,shiraishi}, unlike that in
LSCO~\cite{LSCOGilardi}. It seems very likely from our measurements 
that at high fields, the FLL 
orientation, particularly when the FLL becomes square, is 
caused by the correlation between FLL planes 
and the directions of zeroes of the {\it d}-wave order parameter. 
It would clearly be of great interest to repeat these investigations
in an untwinned crystal, where the effects of any pinning by twin planes
would be completely removed, and the effect of the {\it a/b} anisotropy
on the ``square'' FLL would be crystal-clear. It appears that further investigation of these 
phenomena  will allow stringent tests of theories of the
order parameter in the mixed state of high-$T_{\rm c}$ materials as a
function of angle of field and doping.

This work was performed at the Swiss Spallation Neutron-Source SINQ, Paul Scherrer Institut  (PSI), Villigen, Switzerland.

\bibliographystyle{prsty}

\noindent
$^{\dagger}$\small{e-mail: sp.brown@physics.org}\\
\end{document}